# LEO: An Open-Source Platform for Linking OMERO with Lab Notebooks and Heterogeneous Metadata Sources

Rodrigo Escobar Díaz Guerrero[1,2], Jamile Mohammad Jafari[1,2], Tobias Meyer-Zedler[1,2], Orlando Guntinas-Lichus[3], Michael Schmitt[2], Juergen Popp[1,2], Thomas Bocklitz[1,2].

[1] Leibniz Institute of Photonic Technology Jena, Member of Leibniz research alliance 'Health technologies', Albert-Einstein-Straße 9, 07745, Jena, Germany.

[2] Institute of Physical Chemistry and Abbe Center of Photonics (IPC), Friedrich-Schiller-University, Helmholtzweg 4, 07743, Jena, Germany.

[3]Department of Otorhinolaryngology, Jena University Hospital, Jena, Germany.

In the interdisciplinary field of microscopy research, managing and integrating large volumes of data stored across disparate platforms remains a major challenge. Data types such as bioimages, experimental records, and spectral information are often maintained in separate repositories, each following different management standards. However, linking these data sources across the research lifecycle is essential to align with the FAIR principles of data management: Findability, Accessibility, Interoperability, and Reusability. Despite this need, there is a notable lack of tools capable of effectively integrating and linking data from heterogeneous sources. To address this gap, we present LEO (Linking Electronic Lab Notebooks with OMERO), a web-based platform designed to create and manage links between distributed data systems. LEO was initially developed to link objects between Electronic Lab Notebooks (ELNs) and OMERO (Open Microscopy Environment Remote Objects, but its functionality has since been extended through a plugin-based architecture, allowing the integration of additional data sources. This extensibility makes LEO a scalable and flexible solution for a wide range of microscopy research workflows.

One of the main tasks of any researcher is data management. Research Data Management (RDM) is the process of collecting, storing, sharing, and maintaining data[1]. Good RDM ensures its accuracy, accessibility, and security throughout its lifecycle. However, in microscopy research, data is not only voluminous but also highly heterogeneous[2]. Researchers often rely on a variety of tools to collect experimental records, store raw data, and manage metadata. This fragmentation creates a disconnect between experimental context and the associated imaging data, which can compromise reproducibility and traceability.

Raw imaging data are typically stored in centralized repositories, such as OMERO (Open Microscopy Environment Remote Objects)[3–5], while experimental context, including protocols, annotations, and observations, is recorded separately in Electronic Lab Notebooks (ELNs). This separation often leads to fragmented data management, making it difficult to maintain traceability.

To address this gap, we developed LEO (Linking Electronic Lab Notebooks with OMERO), a cloud-based and modular tool that connects ELNs with OMERO. LEO enables researchers to associate experimental context from lab notebooks with corresponding image datasets stored in OMERO. Built on a flexible plugin-based architecture, LEO can be extended to connect with additional metadata sources beyond ELNs. This integration enhances data traceability by enabling intuitive navigation between experimental records, imaging data, and their associated metadata.

LEO is also aligned with the FAIR principles, which promote data practices that ensure research outputs are Findable, Accessible, Interoperable, and Reusable[6]. By creating persistent links between ELNs and OMERO datasets, enabling standardized metadata exchange, and supporting secure data access, LEO fosters a more transparent and sustainable data ecosystem.

**Software tools**

OMERO (Open Microscopy Environment Remote Objects) is an open-source client-server platform designed for storing, visualizing, annotating, and sharing large and complex biological imaging datasets. It supports a wide variety of microscopy formats and offers powerful tools for managing imaging metadata, organizing datasets, and controlling access in multi-user environments. OMERO has become a key component in many research infrastructures due to its scalability, flexibility, and alignment with FAIR data practices[7,8].

While OMERO addresses the need for structured storage and access to imaging data, it does not directly support the experimental documentation typically recorded in ELNs. Meanwhile, many laboratories still rely on traditional handwritten notebooks, which are difficult to search, share, or back up. Others have adopted digital ELNs, such as eLabFTW, LabArchives, or Labguru, which offer structured, searchable, and often collaborative environments to document research. However, these systems often remain disconnected from data repositories like OMERO, leading to fragmentation between experimental context and the raw imaging data.

Electronic Lab Notebooks (ELNs) have emerged as digital platforms to document experiments, while OMERO serves as a robust system for storing, visualizing, and sharing microscopy data. Despite their strengths, these platforms are rarely integrated, leading to siloed information. Bridging these tools is critical for enhancing data accessibility, reproducibility, and collaboration across research teams.

In this paper, we present the motivation, design, and implementation of LEO. We describe its system architecture, demonstrate its use in practical research settings, and discuss how it contributes to FAIR data practices and improved research workflows.

**Related works**

To our knowledge, there is currently no platform comparable to LEO that simplifies the integration between ELNs and OMERO. However, several alternative approaches have been proposed to address this challenge. For example, Friedmann et al.[9] suggest referencing the unique identification numbers of OMERO objects directly within the ELN. While conceptually straightforward, this method becomes difficult to scale in experiments involving a large number of related images, as it requires manually tracking and inserting numerous identifiers into the ELN.

A similar approach was proposed by Jannasch et al.[10], where ELN entries are linked to OMERO using key-value tables embedded within the OMERO metadata.

Meanwhile, Wagner et al.[11] propose a more sophisticated solution based on a graph database. However, this approach involves collecting data through multiple workflows and tools, making the setup significantly more complex. It demands a high level of user commitment and often requires training to effectively use the entire toolchain.

LEO offers a practical alternative to existing approaches by providing a user-friendly interface that simplifies the linking of metadata across multiple sources. It enables visualization of all connected metadata and streamlines the transfer of experimental information from ELNs into OMERO. Designed to be highly scalable and flexible, LEO is an open-source platform that can be adapted to the diverse needs of research disciplines working with bioimaging data.

# Results

LEO is fully containerized using Docker, enabling easy deployment both on local and on remote servers. We have tested both deployment modes: locally for synthetic testing and in a cloud environment by deploying LEO on the server infrastructure of the Leibniz Institute of Photonic Technology (Leibniz IPHT). In the server deployment,

LEO has been integrated into the workflow of ongoing optics research in combination with life science experiments utilizing OMERO.

The current workflow consists of five main stages:

1) **Experimental stage:** All experiments related to a project are conducted and documented.
2) **Data generation stage:** Data and metadata are collected following the institute's standardized formats and entered into the ELN.
3) **Uploading stage**: The generated images are uploaded to OMERO.
4) **Linking stage:** Experimental records and OMERO images are connected using LEO.
5) **Population stage:** Metadata from the ELNs is transferred and integrated into the corresponding OMERO objects.

## Use cases

### Fluorescence Microscopy Denoising

To demonstrate LEO's utility, we present two representative use cases. The first example uses the Fluorescence Microscopy Denoising (FMD) dataset[12], a publicly available collection of 12,000 real fluorescence microscopy images specifically curated for Poisson-Gaussian denoising tasks. The dataset includes images acquired using three distinct microscopy modalities: confocal, two-photon, and wide-field fluorescence microscopy. These modalities were applied to a variety of biological samples, including zebrafish embryos, mouse brain tissues, and cultured cells. This dataset is a standard dataset for benchmarking image and machine learning algorithms, as performed in the reference[13].

Following our proposal workflow, we began by creating the corresponding entries in the Electronic Lab Notebook (ELN). As this dataset was originally developed outside our institution, we manually curated and extracted relevant information from the original publication to populate the ELNs. A fraction of the ELN can be viewed in Figure 1a.

Then we proceeded to upload the image data to OMERO (Figure 1b). Since this dataset is primarily used to train machine learning models, we chose to upload only the raw and corresponding ground truth images. For organizational purposes, a separate OMERO-dataset object was created for each microscopy modality. Given OMERO's three-tier hierarchical structure (Project, Dataset, and Images), we further annotated the images using tags corresponding to their sample type (e.g., MICE, FISH, and BPAE – bovine pulmonary artery endothelial cells).

Finally, we used LEO to create a link between the OMERO project and the ELN entry. As shown in Figure 1c, LEO provides a summary view of the linked items and also allows users to explore detailed metadata associated with each link. In this particular case, metadata population was not necessary, as the images were not accompanied by sample-specific metadata.

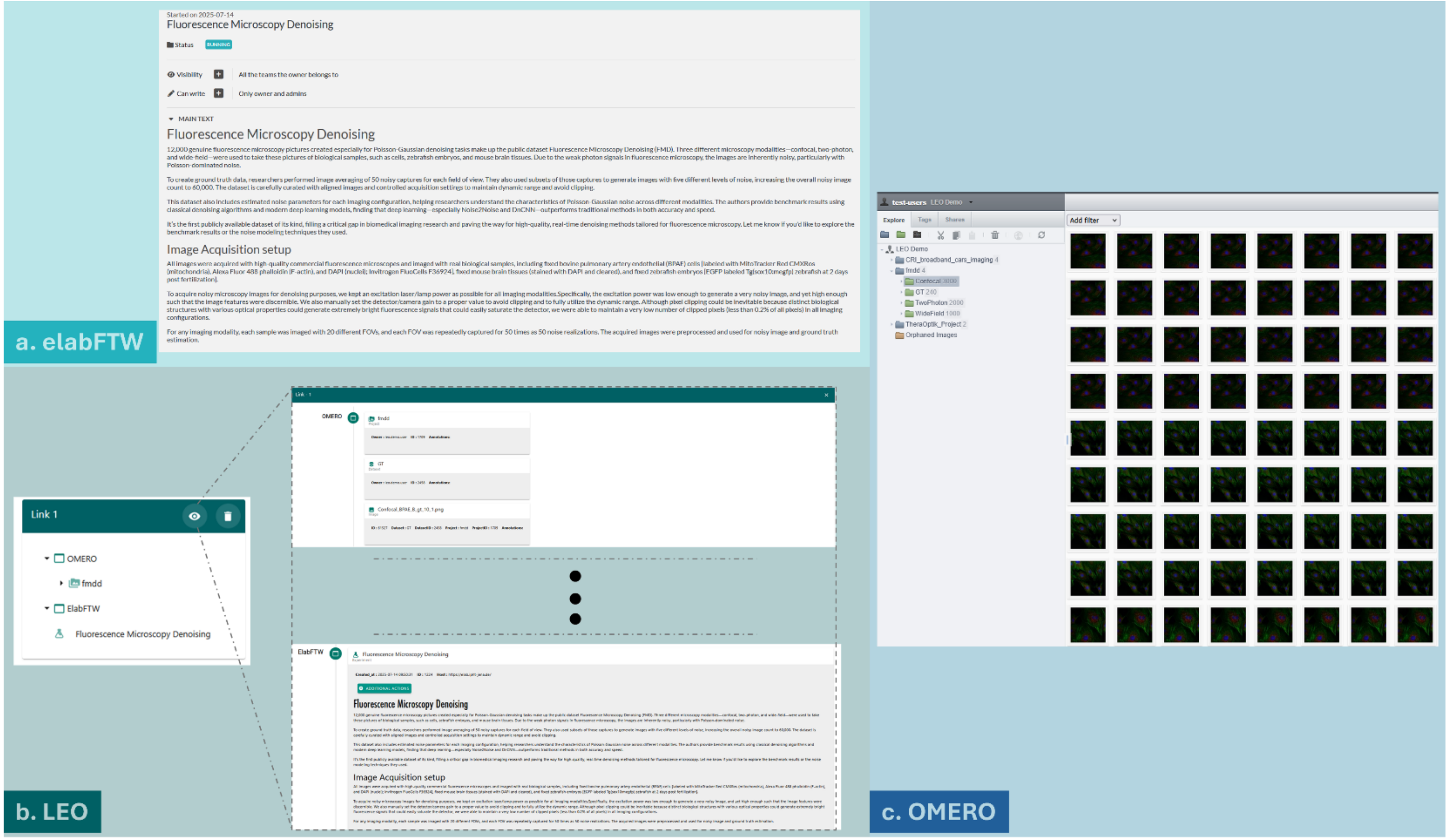

Figure 1.Implementation of the proposed workflow using the Fluorescence Microscopy Denoising (FMD) dataset. Panel **a** shows the main experimental elements recorded in the Electronic Lab Notebook (ELN). Panel **b** displays the link created in LEO, where the left side summarizes the linked elements and the right side presents detailed metadata associated with the ELN and OMERO objects. Panel **c** shows the dataset uploaded to OMERO, organized by imaging modality.

**TheraOptik dataset**

The second use case involves a dataset generated in the TheraOptik project[14,15]. The goal of this project was to develop a fiber-based measurement system that integrates multimodal nonlinear microscopy and femtosecond (fs) laser ablation[16]. Specifically, we showcase a portion of the data used in[14] . This dataset was created for a preclinical study aimed at diagnosing head and neck cancer using multimodal nonlinear optical imaging combined with deep learning based semantic segmentation.

The dataset comprises twenty-three multimodal nonlinear images, each containing three channels, with each channel representing a different imaging modality: CARS (coherent anti-Stokes Raman scattering) in the red channel, TPEF (two-photon excited fluorescence) in the green channel, and SHG (second-harmonic generation) in the blue channel.

For this example, we selected the raw images and the illumination-corrected images. The correction was applied to address uneven illumination artifacts using the BaSiC algorithm[17] implemented in Fiji/ImageJ[18]. An example is shown in Supplementary Figure 1, where the raw and corrected versions of the same image are presented side by side.

Following our proposed workflow, we first created a new experiment entry in the ELN using elabFTW. The metadata of this dataset is publicly available, so it is possible to access complete experimental records and metadata.

We organized the metadata into two structured tables: One for patient-related information (e.g., age, gender) and another for optical measurement parameters (e.g., number of tiles, scanning amplitude, Stokes power). The metadata are shown in Supplementary Table I (patient metadata) and Supplementary Table II (optical measurements).

Next, we uploaded the images to OMERO, organizing them into two separate OMERO-datasets: one containing the raw images, and another with the corrected versions. We then used LEO to link the ELN experiment to both OMERO datasets.

Finally, we used LEO's metadata population tools (described in detail in later sections) to transfer the structured metadata from the ELN into OMERO. After this step, the metadata became accessible and viewable both in LEO and OMERO. As illustrated in Figure 2, the left panels show the metadata views before population, while the right panels show the enriched views after the metadata was transferred.

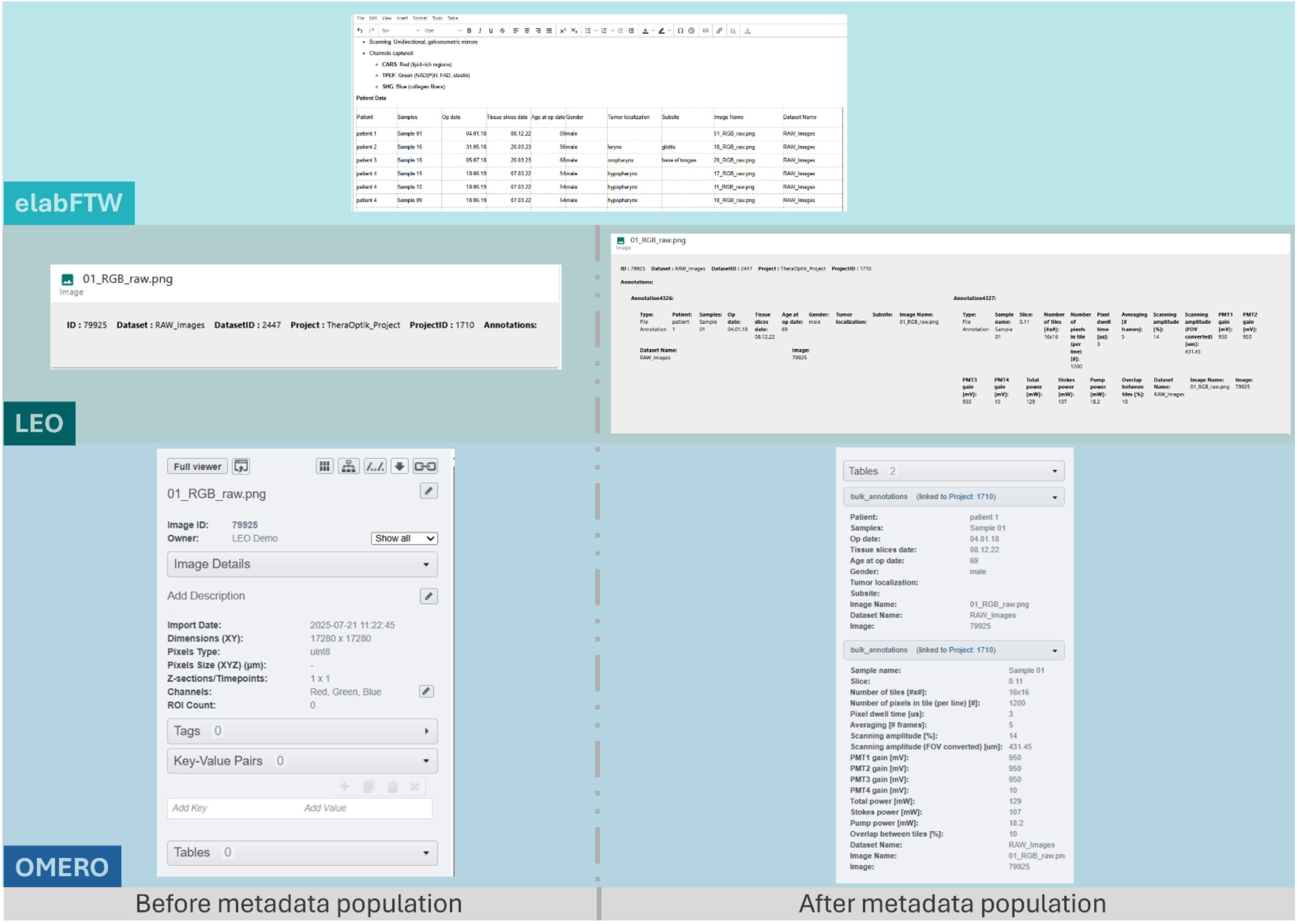

Figure 2.Metadata views for one image in LEO (middle), OMERO (bottom), and the original metadata from the experiment ELN (top). Left: before population; Right: after metadata has been added from the ELN.

# Discussion

In this work, we introduced LEO, a web-based platform developed to address a pressing need in microscopy research: the integration of experimental records and imaging data from disparate systems in a FAIR-compliant manner. By linking ELNs with imaging repositories such as OMERO, LEO offers a user-friendly interface and a plugin-based architecture that fosters flexibility, interoperability, and scalability across research environments.

Unlike other approaches that rely on manual referencing or complex toolchains, LEO enables a linking of metadata and imaging objects through a user-friendly interface. It also supports the direct population of OMERO metadata using structured tables from ELN entries, greatly simplifying a task that typically requires scripting or specialized tools. Through its plugin system, LEO can be extended to accommodate additional data providers beyond ELNs and OMERO.

We demonstrated the platform's applicability by deploying it both locally for testing purposes and on the production server at the Leibniz Institute of Photonic Technology, where it has been integrated into active research workflows involving optics and life sciences. The five-stage research process (experiment, data generation, uploading, linking, and population) highlights LEO's ability to streamline data handling and enhance the traceability and reusability of research outputs.

To further illustrate its versatility, we applied LEO in two representative use cases. The first involved a publicly available fluorescence microscopy dataset used in algorithm development for image denoising, while the second showcased a preclinical multimodal imaging study. These examples demonstrate LEO's flexibility in supporting metadata integration from external datasets and experimental workflows within ongoing research.

Looking ahead, we plan to expand LEO's functionality by incorporating support for additional data modalities (e.g., spectral metadata) and integrating a robust search engine. We also aim to foster community contributions to the plugin ecosystem, enabling LEO to serve as a shared infrastructure for research data management across institutions and disciplines.

In summary, LEO provides a practical and extensible solution for integrating experimental metadata with imaging data, helping researchers move toward more reproducible, transparent, and FAIR-aligned scientific practices.

# Methods

LEO is a cloud-based solution, developed using the Django framework (v4.2) for the backend. For the frontend, the Vue.js JavaScript framework (v3.4) was used in combination with the Vuetify UI library (v3.5.17). LEO was designed with flexibility and scalability in mind. Its plugin-based architecture allows users to connect multiple types of data providers. These providers currently include ELNs and OMERO server instances, and the system is extensible to support future integrations, such as spectral metadata repositories or additional data platforms.

Two primary workflows define how users interact with LEO (Figure 3):

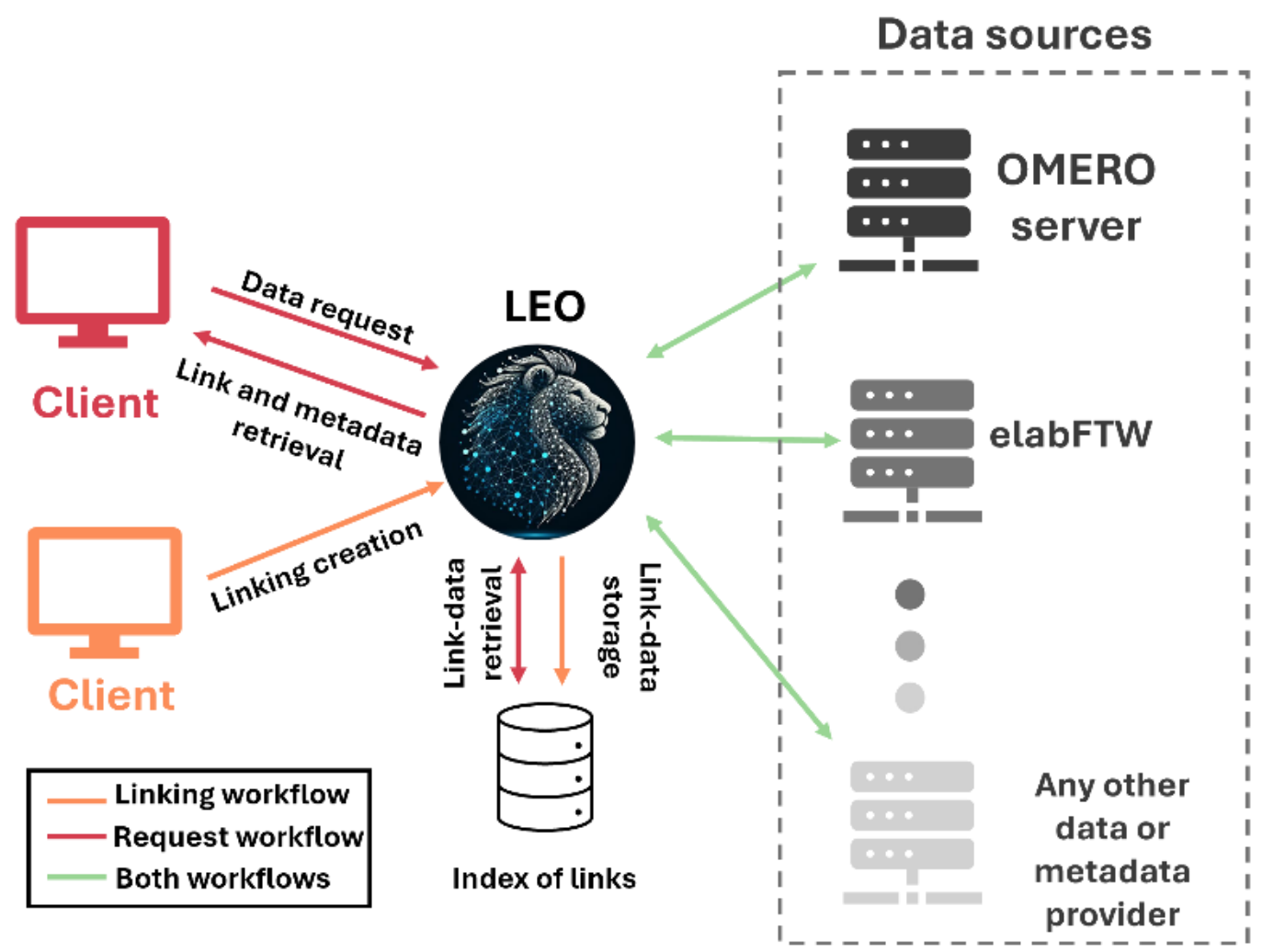

Figure 3. LEO workflows schema. The link workflow is shown in orange, the request workflow in red, and the paths used for both workflows are in green.

1. **Linking Workflow:** This workflow enables users to browse and select objects from any instance of a data provider and establish links between them. For example, a user can link a specific ELN experiment to a corresponding OMERO object (such as datasets, images, or projects).
2. **Request Workflow:** This workflow allows users to retrieve and visualize existing links, including metadata from the connected instance, from data providers.

LEO consists of four core components: an authentication system, a visualization system, a plugin system, and a metadata population system. These components work together to ensure secure access, facilitate linked data visualization, support integration with various data providers, and handle the extraction and linking of metadata across connected systems.

### Authentication system

To manage access to the platform, a custom authentication system was implemented instead of Django's default authentication mechanism. When a user attempts to log in, LEO establishes a connection with the OMERO server, which validates the user's credentials. If the authentication is successful, LEO creates a new session using these credentials. Then, the session credentials are encrypted using a 32-byte key derived from the SHA-256 hash of Django's SECRET_KEY parameter, the session is open until the user finalize the session.

This authentication approach offers several advantages:

1) Users who already have an OMERO account do not need to remember an additional username and password.
2) All credential validation and protection are delegated to OMERO—a mature, widely adopted platform with robust security practices. This minimizes security risks and leverages OMERO's well-established infrastructure for user management.

## Visualization system

LEO features three main tabs. The first is the Settings tab, which allows users to create and manage instances of data providers based on the available plugins (the plugin system will be described in a later section). For example, consider a scenario where you are collaborating on a project between two institutions, both using the same ELN, such as elabFTW, Chemotion ELN, or Kadi4Mat to register experimental data, but each hosted on a different domain. In this case, LEO allows users to connect to both "Laboratory A" and "Laboratory B" using the same plugin, since they share a common ELN backend. Alternatively, if each laboratory uses a different ELN, LEO can support both through separate plugins, one for each system. Each plugin is responsible for handling the connection setup and for retrieving metadata in a standardized format, ensuring interoperability regardless of the ELN or its hosting domain. Additional plugins can be developed and added to extend compatibility with other ELNs or data systems as needed.

The main elements of the Settings tab are shown in Figure 4a. To create a new instance of a data provider, the user simply clicks the "+ ADD" button. This opens a form where only three fields need to be completed: (1) the host domain, (2) an API key (if required by the data provider), and (3) the data provider name, selected from those defined in the plugin system. Once the form is submitted and the data is validated, the configuration is saved.

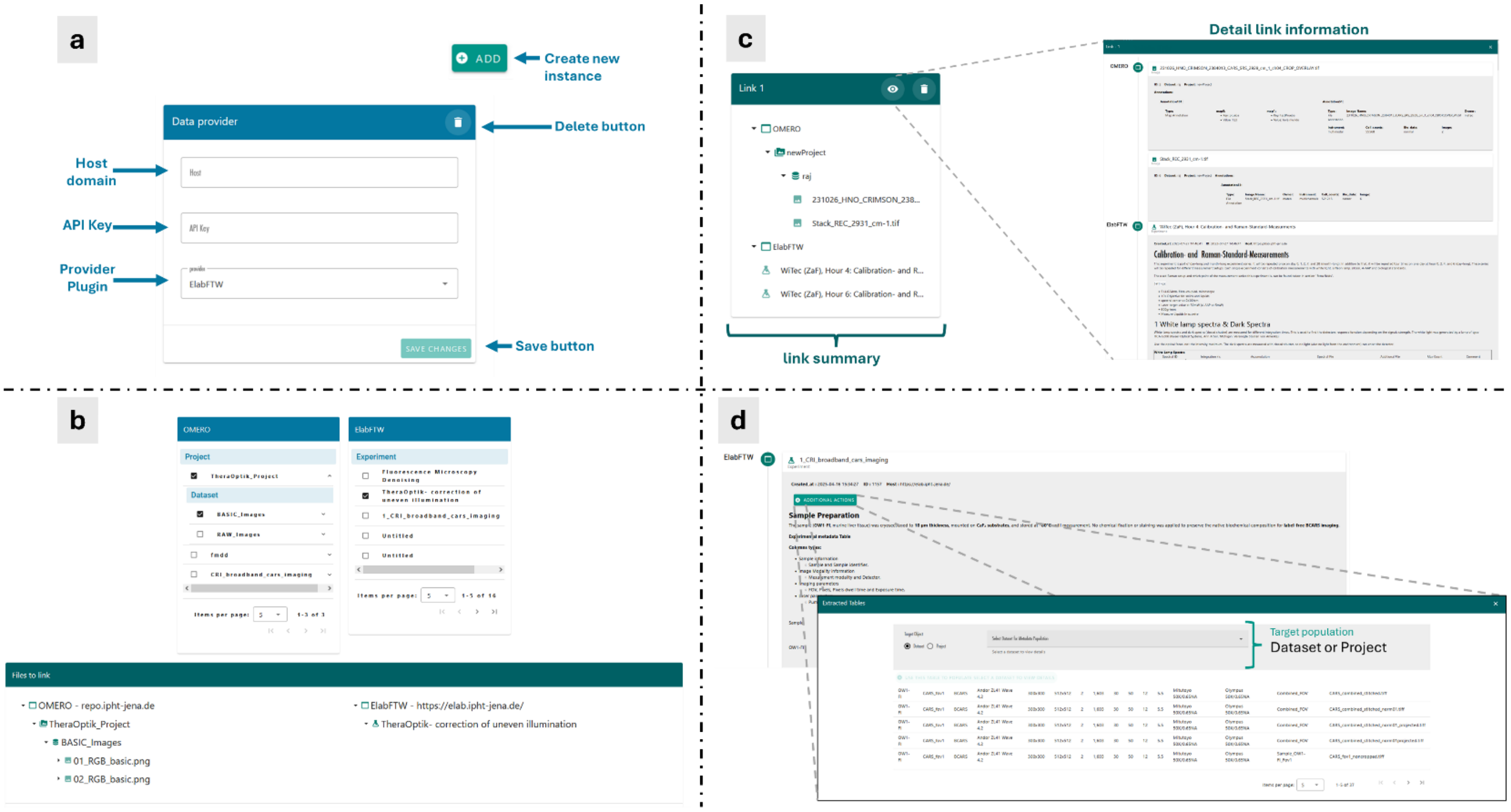

Figure 4. LEO user interface. Panel **a** shows the main components of the *Settings* tab, where users can configure instances of various data providers. Panel **b** illustrates the *Link Creation* tab: the top (blue) section displays available objects from any configured data provider, while the bottom (green) section shows the selected items ready to be linked. Panel **c** presents an example link within the *Current Links* tab, with a summary on the left and detailed link information on the right. Panel **d** demonstrates the metadata population feature; by clicking the "Additional Actions" button, users can extract all tables embedded in an ELN experiment and use them to populate metadata for linked OMERO objects.

The second tab is Link Creation (Figure 4b), which, as the name suggests, is where users create links between data provider instances. Once the provider instances have been properly configured, the window displays a panel for each instance. Each panel shows the available objects that can be linked—for example, in OMERO, users can link images, datasets, and projects, while in eLabFTW, the linkable objects are experiments.

Users can select the objects they want to link using checkboxes. A preview section at the bottom of the interface displays the selected elements. After verifying the selection, the links can be created by clicking the Save button.

The final tab, Current Links (Figure 4c), displays all previously created links. Each link includes a summarized view of the connected elements, allowing users to quickly identify the linked resources. Additionally, users can expand each entry to view more detailed information, such as the full description of an experiment from an ELN or all associated metadata for an image stored in OMERO.

**Metadata population system**

While the primary goal of LEO is to establish links between objects and metadata from different sources, we have also implemented an additional feature: the ability to populate metadata from an ELN experiment into OMERO. It is a common practice to use tables within electronic lab notebooks to record essential experimental information, such as biological sample characteristics, optical configurations, or instrument settings, rather than storing this data in separate databases. Although OMERO offers several mechanisms to store metadata, these are often not user-friendly or easily accessible to non-experts. To address this gap, LEO introduces a simplified metadata population system that allows researchers to transfer tabular metadata from ELNs directly into OMERO.

This functionality is based on and shares characteristics with the omero-metadata Python library[19], which is widely used for structured metadata handling in OMERO. In LEO, metadata population is initiated by navigating to the detailed view of a linked experiment and clicking the “Additional Actions” button. This action opens a new window that automatically detects and displays all tables embedded in the body of the experiment. Users can then select which table to use and choose a target OMERO object (either a dataset or a project) from a dropdown list that includes only the objects already linked to the experiment. An example view of this functionality is shown in Figure 4d.

To ensure correct metadata mapping, the tables must follow the following considerations:

- The first row of the table must contain column headers.
- Users may include as many columns as needed, using any header names.
- If the target object is a project, the table must include both a "Dataset Name" column and an "Image Name" column.

- If the target is a dataset, only the "Image Name" column is required.
- The values in these columns must exactly match the names of the corresponding datasets and images in OMERO.
- If a table is used to populate an object that already contains a previous table with the same headers, the existing table will be overwritten and replaced with the new one.

**Plugin System**

One of LEO's key features is its extensible plugin system, which allows additional data providers to be integrated in the future. This design gives LEO excellent scalability and adaptability, making it suitable for a wide range of research environments and user needs.

LEO implements a discovery and registration mechanism for plugins (Figure 5). When the platform starts, it automatically scans for installed plugins. All detected plugins are registered and made available to the user.

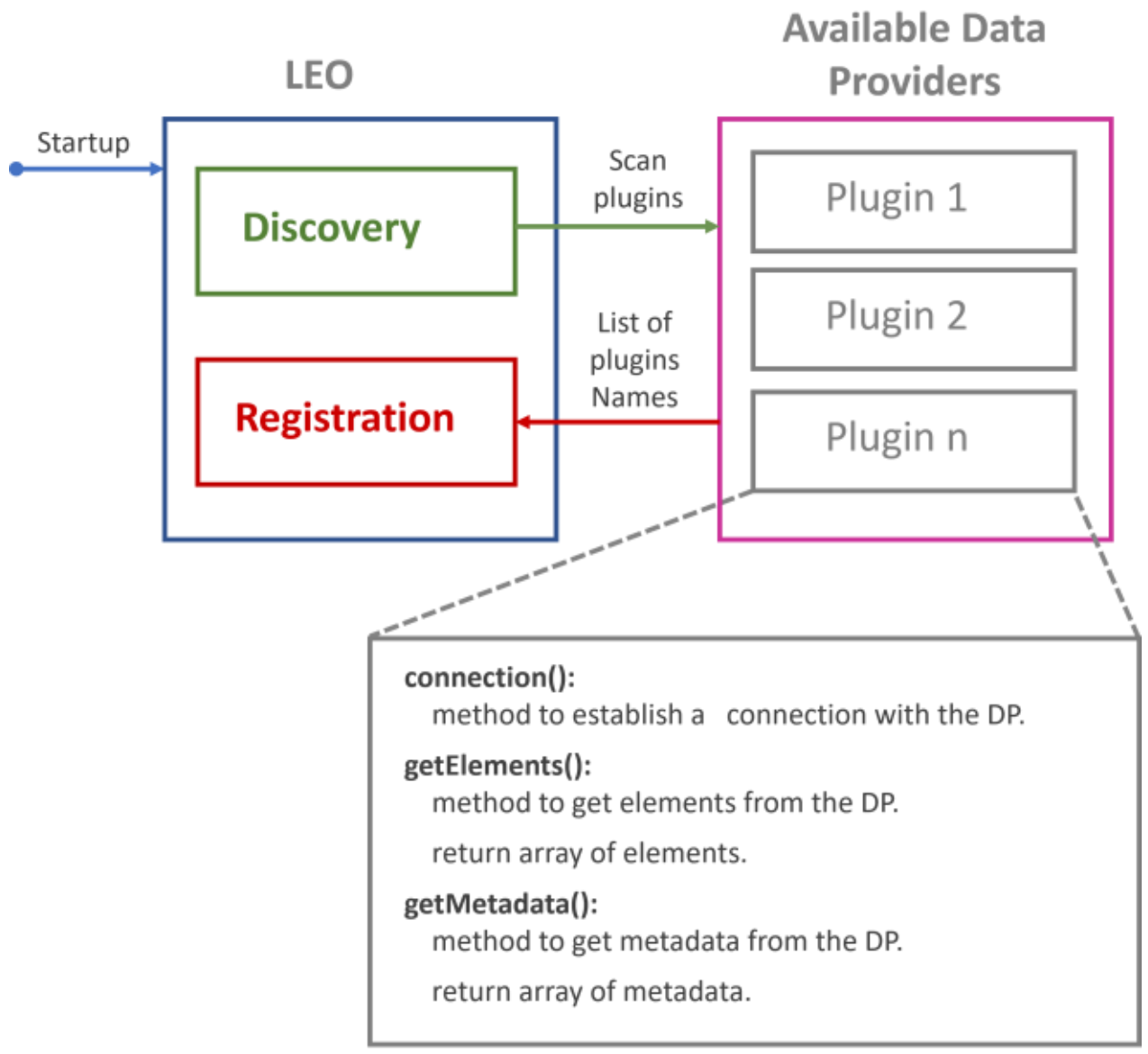

Figure 6. Plugin system architecture. The plugin system uses a discovery and registration mechanism. When the platform starts, it automatically scans for installed plugins. All detected plugins are registered and made available to the user.

To support integration with new data providers, LEO uses a standardized plugin interface. Each plugin must inherit from a core class and implement specific methods. The following pseudocode shows the structure required to develop a new plugin:

Structure of a new data provider plugin

```
Input:
    request = request object from user session
    host = domain/URL of the data provider
    elements = array of items selected from the data provider
Output:
    arrayOfElements = elements available for linking
    arrayOfMetadata = metadata associated with selected elements

Function connection (request, host):
    // Apply connection logic to the host
    return connection

Function getElements (request):
    connection = connection(request, host)
    // Retrieve and return array of elements
    return arrayOfElements = [
        {origin_id: string, unique ID in the data provider,
        title: string, name of the element,
        type: string, e.g., "Experiment", "Image", etc. ,
        id: integer, unique in this request,
        // optional attributes
        children: array of nested elements,
        parent: string (parent ID) or "false" if none}, ...]

Function getMetadata(request, elements):
    connection = connection(request, host)
    // Retrieve and return metadata for selected elements
    return arrayOfMetadata = [
        {Type: string, type of element e.g., "Experiment", "Image", etc. ,
        Name: string, name of element,
        ID: string, internal ID in data provider,
        // optional: any extra keys (plain or HTML-formatted)
        }, ...]
```

Figure 5.Pseudocode for the implementation of a new data provider plugin.

As is shown in the pseudocode of Figure 6, creating a new plugin in LEO requires defining a single class that implements three main methods: one to establish a connection with the data provider, another to retrieve the elements available for linking (e.g., experiments, images, projects), and a third to extract the associated metadata. Plugins use Python dictionaries with a key:value structure, allowing a high degree of flexibility. The number and types of metadata fields are not fixed, so developers can include as much detail as needed.

The steps to implement a new plugin are straightforward:

1) Develop and test the plugin independently, outside of LEO.
2) Add the plugin code to the appropriate plugins folder within the LEO project.
3) Register the plugin name in the database as a new data provider.
4) Restart LEO to activate the new plugin.

# Data Availability

The Fluorescence Microscopy Denoising (FMD) dataset is publicly available at the reference[20]. The TheraOptik dataset is also publicly available at the reference[15].

# Code Availability

LEO is open-source software and is available on GitHub at the reference[21].

# References

1. Perrier, L. *et al.* Research data management in academic institutions: A scoping review. *PLoS One* **12**, e0178261 (2017).

2. Wallace, C. T., St. Croix, C. M. & Watkins, S. C. Data management and archiving in a large microscopy-and-imaging, multi-user facility: Problems and solutions. *Mol Reprod Dev* **82**, 630–634 (2015).

3. Li, S. *et al.* Metadata management for high content screening in OMERO. *Methods* **96**, 27–32 (2016).

4. Burel, J.-M. *et al.* Publishing and sharing multi-dimensional image data with OMERO. *Mammalian Genome* **26**, 441–447 (2015).

5. Williams, E. *et al.* Image Data Resource: a bioimage data integration and publication platform. *Nat Methods* **14**, 775–781 (2017).

6. Wilkinson, M. D. *et al.* The FAIR Guiding Principles for scientific data management and stewardship. *Sci Data* **3**, 160018 (2016).

7. Allan, C. *et al.* OMERO: flexible, model-driven data management for experimental biology. *Nat Methods* **9**, 245–253 (2012).

8. OMERO | Open Microscopy Environment (OME). https://www.openmicroscopy.org/omero/.

9. Friedmann, V., Bellmer, M., Antons, R. & Hafok, H. Automated quantitative microstructural analysis using the open-source server-client platform OMERO. *Practical Metallography* **60**, 820–837 (2023).

10. Jannasch, A. *et al.* Setting up an institutional OMERO environment for bioimage data: Perspectives from both facility staff and users. *J Microsc* **297**, 105–119 (2025).

11. Wagner, R., Waltemath, D., Yordanova, K. & Becker, M. M. Towards FAIR Data Workflows for Multidisciplinary Science: Ongoing Endeavors and Future Perspectives in Plasma Technology. (2024) doi:10.5220/0012808000003756.

12. Zhang, Y. *et al.* A Poisson-Gaussian Denoising Dataset With Real Fluorescence Microscopy Images. in *2019 IEEE/CVF Conference on Computer Vision and Pattern Recognition (CVPR)* 11702–11710 (IEEE, 2019). doi:10.1109/CVPR.2019.01198.

13. Corbetta, E. & Bocklitz, T. Machine Learning-Based Estimation of Experimental Artifacts and Image Quality in Fluorescence Microscopy. *Advanced Intelligent Systems* **7**, (2025).

14. Calvarese, M. *et al.* Endomicroscopic AI-driven morphochemical imaging and fs-laser ablation for selective tumor identification and selective tissue removal. *Sci Adv* **10**, 9721 (2024).

15. Calvarese, M. *et al.* supporting_data_for_Endomicroscopic_AI-driven_morphochemical_imaging_and_fs-laser_ablation_for_selective_tumor_identification_and_selective_tissue_removal. in *Dataset on Zenodo* doi:10.5281/zenodo.14604803.

16. TheraOptik - Leibniz-Institut für Photonische Technologien e.V. https://www.leibniz-ipht.de/de/forschung/projekte/theraoptik/.

17. Peng, T. *et al.* A BaSiC tool for background and shading correction of optical microscopy images. *Nat Commun* **8**, 14836 (2017).

18. Schindelin, J. *et al.* Fiji: an open-source platform for biological-image analysis. *Nat Methods* **9**, 676–682 (2012).

19. omero-metadata·PyPI. https://pypi.org/project/omero-metadata/.

20. Fluorescence Microscopy Denoising (FMD) dataset. https://curate.nd.edu/articles/dataset/Fluorescence_Microscopy_Denoising_FMD_dataset/24744648.

21. GitHub - NFDI4BIOIMAGE/LEO: Linking Electronic Lab Notebooks and other sources with OMERO objects. https://github.com/NFDI4BIOIMAGE/LEO.

# Author Contributions

TB conceived the research project, contributed to its conceptualization, provided project supervision, and conducted in-depth reviews that improved the software's quality and functionality. JJ supported the conceptualization phase and was involved in software testing. OG, TM, MS, and JP contributed to the acquisition of the TheraOptik dataset and provided conceptual input that informed improvements to the software at various stages of development, and reviewed the final version of the manuscript. RE conceived the original software design, implemented the software, and wrote the first draft of the manuscript.

# Competing Interests

All the authors declare the absence of any competing interests.

# Acknowledgements

Funded by the Deutsche Forschungsgemeinschaft (DFG, German Research Foundation) – 501864659 (NFDI4Bioimage) and co-funded by the European Union (ERC, STAIN-IT, 101088997). Views and opinions expressed are however those of the author(s) only and do not necessarily reflect those of the European Union or the European Research Council. Neither the European Union nor the granting authority can be held responsible for them. We also thank the authors of reference [12] and [14] for data acquisition.

# Supplementary Information

## LEO: An Open-Source Platform for Linking OMERO with Lab Notebooks and Heterogeneous Metadata Sources

Rodrigo Escobar Díaz Guerrero[1,2], Jamile Mohammad Jafari[1,2], Tobias Meyer-Zedler[1,2], Orlando Guntinas-Lichus[3], Michael Schmitt[2], Juergen Popp[1,2], Thomas Bocklitz[1,2].

[1] Leibniz Institute of Photonic Technology Jena, Member of Leibniz research alliance 'Health technologies', Albert-Einstein-Straße 9, 07745, Jena, Germany.

[2] Institute of Physical Chemistry and Abbe Center of Photonics (IPC), Friedrich-Schiller-University, Helmholtzweg 4, 07743, Jena, Germany.

[3]Department of Otorhinolaryngology, Jena University Hospital, Jena, Germany.

# Contents

# Supplementary Figures

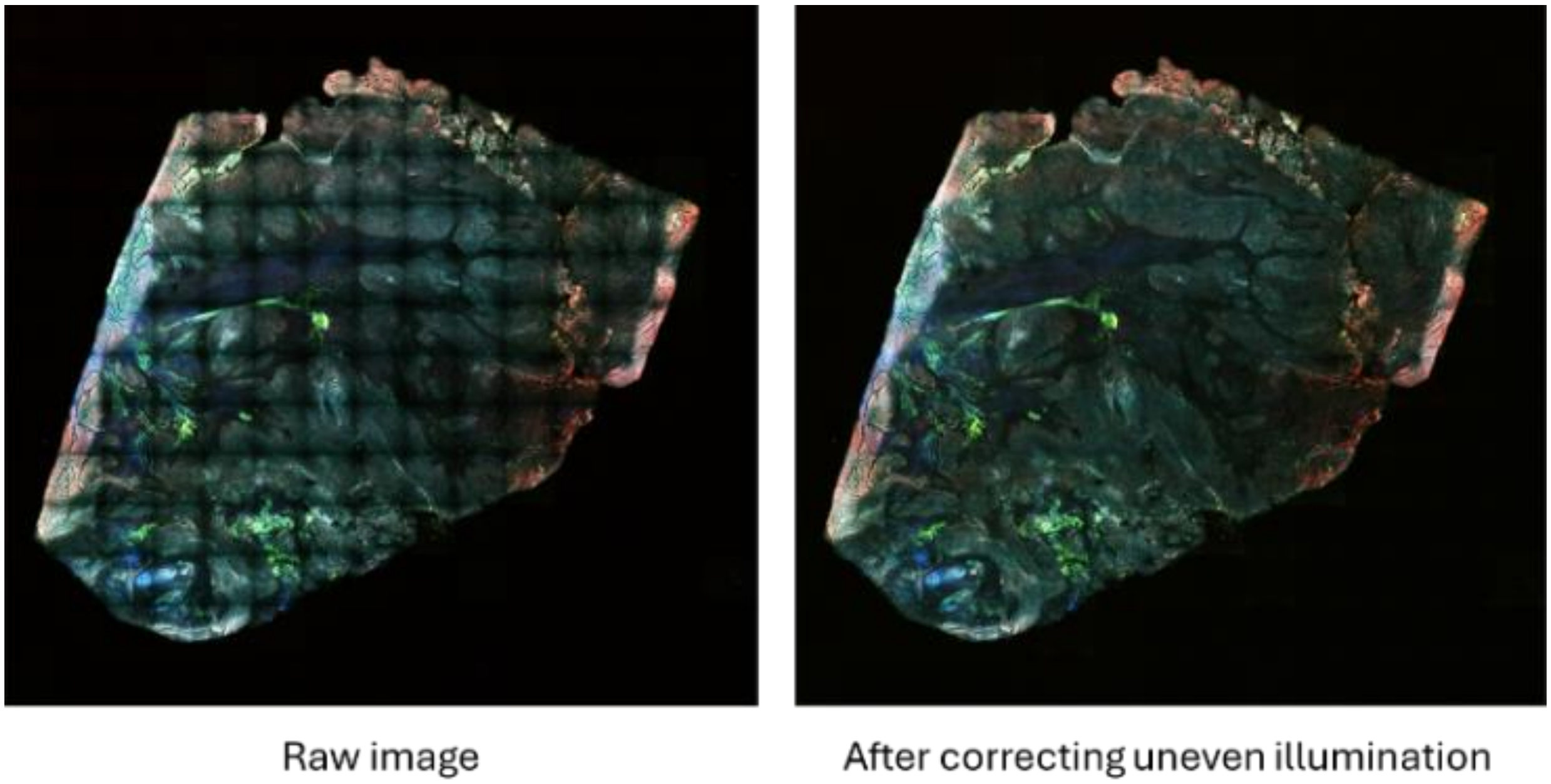

Supplementary Figure 1. Sample images from the TheraOptic dataset: on the left, a raw image, and on the right, the same image after illumination correction using BaSiC in Fiji.

# Supplementary Tables

*Supplementary Table 1. Patient metadata for the TheraOptic project. Op =Operation.*

| Patient | Samples | Op date | Tissue slices date | Age at op date (in years) | Gender | Tumor localization | Subsite | Image Name | Dataset Name |
|---|---|---|---|---|---|---|---|---|---|
| patient 1 | Sample 01 | 04.01.18 | 08.12.22 | 69 | male | | | 01_RGB_raw.png | RAW_Images |
| patient 2 | Sample 16 | 31.05.18 | 20.03.23 | 56 | male | larynx | glottis | 18_RGB_raw.png | RAW_Images |
| patient 3 | Sample 18 | 05.07.18 | 20.03.23 | 68 | male | oropharynx | base of tongue | 20_RGB_raw.png | RAW_Images |
| patient 4 | Sample 15 | 18.06.19 | 07.03.22 | 54 | male | hypopharynx | | 17_RGB_raw.png | RAW_Images |
| patient 4 | Sample 10 | 18.06.19 | 07.03.22 | 54 | male | hypopharynx | | 11_RGB_raw.png | RAW_Images |
| patient 4 | Sample 09 | 18.06.19 | 07.03.22 | 54 | male | hypopharynx | | 10_RGB_raw.png | RAW_Images |
| patient 4 | Sample 19 | 18.06.19 | 20.03.23 | 54 | male | hypopharynx | | 21_RGB_raw.png | RAW_Images |
| patient 4 | Sample 19 | 18.06.19 | 20.03.23 | 54 | male | hypopharynx | | 22_RGB_raw.png | RAW_Images |
| patient 4 | Sample 20 | 04.07.19 | 20.03.23 | 61 | male | hypopharynx | | 23_RGB_raw.png | RAW_Images |
| patient 5 | Sample 17 | 04.07.19 | 20.03.23 | 61 | male | hypopharynx | | 19_RGB_raw.png | RAW_Images |
| patient 6 | Sample 14 | 08.07.19 | 07.03.22 | 67 | male | hypopharynx | | 15_RGB_raw.png | RAW_Images |
| patient 6 | Sample 14 | 08.07.19 | 07.03.22 | 67 | male | hypopharynx | | 16_RGB_raw.png | RAW_Images |
| patient 7 | Sample 05 | 23.07.19 | 07.03.22 | 61 | male | hypopharynx | | 06_RGB_raw.png | RAW_Images |
| patient 7 | Sample 07 | 08.08.19 | 07.03.22 | 61 | male | hypopharynx | | 08_RGB_raw.png | RAW_Images |
| patient 8 | Sample 11 | 16.09.19 | 07.03.22 | 64 | male | larynx | epiglottis | 12_RGB_raw.png | RAW_Images |
| patient 9 | Sample 03 | 18.03.20 | 07.03.22 | 66 | male | oral cavity | floor of mouth | 04_RGB_raw.png | RAW_Images |
| patient 10 | Sample 06 | 19.03.20 | 07.03.22 | 34 | female | oral cavity | tongue/floor of mouth | 07_RGB_raw.png | RAW_Images |
| patient 11 | Sample 04 | 31.03.20 | 07.03.22 | 47 | male | oral cavity | | 05_RGB_raw.png | RAW_Images |
| patient 12 | Sample 02 | 30.07.20 | 08.12.22 | 77 | male | oral cavity | tongue | 02_RGB_raw.png | RAW_Images |
| patient 12 | Sample 02 | 30.07.20 | 08.12.22 | 77 | male | oral cavity | tongue | 03_RGB_raw.png | RAW_Images |
| patient 13 | Sample 08 | 29.04.22 | 07.03.22 | 60 | male | oral cavity | floor of mouth | 09_RGB_raw.png | RAW_Images |
| patient 14 | Sample 12 | 25.01.18 | 20.03.23 | 53 | male | oropharynx | tonsil | 13_RGB_raw.png | RAW_Images |
| patient 15 | Sample 13 | 29.11.18 | 20.03.23 | 58 | male | larynx | glottis | 14_RGB_raw.png | RAW_Images |
| patient 1 | Sample 01 | 04.01.18 | 08.12.22 | 69 | male | | | 01_RGB_basic.png | BASIC_Images |
| patient 2 | Sample 16 | 31.05.18 | 20.03.23 | 56 | male | larynx | glottis | 18_RGB_basic.png | BASIC_Images |
| patient 3 | Sample 18 | 05.07.18 | 20.03.23 | 68 | male | oropharynx | base of tongue | 20_RGB_basic.png | BASIC_Images |
| patient 4 | Sample 15 | 18.06.19 | 07.03.22 | 54 | male | hypopharynx | | 17_RGB_basic.png | BASIC_Images |
| patient 4 | Sample 10 | 18.06.19 | 07.03.22 | 54 | male | hypopharynx | | 11_RGB_basic.png | BASIC_Images |
| patient 4 | Sample 09 | 18.06.19 | 07.03.22 | 54 | male | hypopharynx | | 10_RGB_basic.png | BASIC_Images |
| patient 4 | Sample 19 | 18.06.19 | 20.03.23 | 54 | male | hypopharynx | | 21_RGB_basic.png | BASIC_Images |
| patient 4 | Sample 19 | 18.06.19 | 20.03.23 | 54 | male | hypopharynx | | 22_RGB_basic.png | BASIC_Images |
| patient 4 | Sample 20 | 04.07.19 | 20.03.23 | 61 | male | hypopharynx | | 23_RGB_basic.png | BASIC_Images |
| patient 5 | Sample 17 | 04.07.19 | 20.03.23 | 61 | male | hypopharynx | | 19_RGB_basic.png | BASIC_Images |
| patient 6 | Sample 14 | 08.07.19 | 07.03.22 | 67 | male | hypopharynx | | 15_RGB_basic.png | BASIC_Images |
| patient 6 | Sample 14 | 08.07.19 | 07.03.22 | 67 | male | hypopharynx | | 16_RGB_basic.png | BASIC_Images |
| patient 7 | Sample 05 | 23.07.19 | 07.03.22 | 61 | male | hypopharynx | | 06_RGB_basic.png | BASIC_Images |
| patient 7 | Sample 07 | 08.08.19 | 07.03.22 | 61 | male | hypopharynx | | 08_RGB_basic.png | BASIC_Images |
| patient 8 | Sample 11 | 16.09.19 | 07.03.22 | 64 | male | larynx | epiglottis | 12_RGB_basic.png | BASIC_Images |
| patient 9 | Sample 03 | 18.03.20 | 07.03.22 | 66 | male | oral cavity | floor of mouth | 04_RGB_basic.png | BASIC_Images |
| patient 10 | Sample 06 | 19.03.20 | 07.03.22 | 34 | female | oral cavity | tongue/floor of mouth | 07_RGB_basic.png | BASIC_Images |
| patient 11 | Sample 04 | 31.03.20 | 07.03.22 | 47 | male | oral cavity | | 05_RGB_basic.png | BASIC_Images |
| patient 12 | Sample 02 | 30.07.20 | 08.12.22 | 77 | male | oral cavity | tongue | 02_RGB_basic.png | BASIC_Images |
| patient 12 | Sample 02 | 30.07.20 | 08.12.22 | 77 | male | oral cavity | tongue | 03_RGB_basic.png | BASIC_Images |
| patient 13 | Sample 08 | 29.04.22 | 07.03.22 | 60 | male | oral cavity | floor of mouth | 09_RGB_basic.png | BASIC_Images |
| patient 14 | Sample 12 | 25.01.18 | 20.03.23 | 53 | male | oropharynx | tonsil | 13_RGB_basic.png | BASIC_Images |
| patient 15 | Sample 13 | 29.11.18 | 20.03.23 | 58 | male | larynx | glottis | 14_RGB_basic.png | BASIC_Images |

*Supplementary Table 2. Optical measurement metadata for the TheraOptic project.*

| Sample name | Slice | Number of tiles | Number of pixel in tile (per line) | Pixel dwell time [us] | Averaging [# frames] | Scanning amplitude [%] | Scanning amplitude [um] | PMT1 gain [mV] | PMT2 gain [mV] | PMT3 gain [mV] | PMT4 gain [mV] | Total power [mW] | Stokes power [mW] | Pump power [mW] | Overlap between tiles [%] | Dataset Name | Image Name |
|---|---|---|---|---|---|---|---|---|---|---|---|---|---|---|---|---|---|
| Sample 01 | 0.11 | 16x16 | 1200 | 3 | 5 | 14 | 431.45 | 950 | 950 | 950 | 10 | 129 | 107 | 18.2 | 10 | RAW_Images | 01_RGB_raw.png |
| Sample 02 | 0.11 | 10x11 | 1200 | 3 | 5 | 14 | 431.45 | 950 | 950 | 950 | 10 | 129 | 107 | 18.2 | 10 | RAW_Images | 02_RGB_raw.png |
| Sample 02 | 0.11 | 12x12 | 1200 | 3 | 5 | 14 | 431.45 | 950 | 950 | 950 | 10 | 129 | 107 | 18.2 | 10 | RAW_Images | 03_RGB_raw.png |
| Sample 03 | 0.11 | 10x11 | 1200 | 3 | 5 | 14 | 431.45 | 950 | 950 | 950 | 10 | 128 | 107 | 17.8 | 10 | RAW_Images | 04_RGB_raw.png |
| Sample 04 | 0.11 | 14x16 | 1200 | 3 | 5 | 14 | 431.45 | 950 | 950 | 950 | 10 | 128 | 107 | 17.8 | 10 | RAW_Images | 05_RGB_raw.png |
| Sample 05 | 0.11 | 8x14 | 1200 | 3 | 5 | 14 | 431.45 | 950 | 950 | 950 | 10 | 129 | 106 | 18.5 | 10 | RAW_Images | 06_RGB_raw.png |
| Sample 06 | 0.11 | 16x16 | 1200 | 3 | 5 | 14 | 431.45 | 950 | 950 | 950 | 10 | 129 | 106 | 18.5 | 10 | RAW_Images | 07_RGB_raw.png |
| Sample 07 | 0.11 | 14x10 | 1200 | 3 | 5 | 14 | 431.45 | 950 | 950 | 950 | 10 | 129 | 107 | 18.1 | 10 | RAW_Images | 08_RGB_raw.png |
| Sample 08 | 0.11 | 14x12 | 1200 | 3 | 5 | 14 | 431.45 | 950 | 950 | 950 | 10 | 129 | 107 | 18.1 | 10 | RAW_Images | 09_RGB_raw.png |
| Sample 09 | 0.11 | 15x16 | 1200 | 3 | 5 | 14 | 431.45 | 950 | 950 | 950 | 10 | 129 | 107 | 17.7 | 10 | RAW_Images | 10_RGB_raw.png |
| Sample 10 | 0.11 | 15x14 | 1200 | 3 | 5 | 14 | 431.45 | 950 | 950 | 950 | 10 | 129 | 107 | 17.7 | 10 | RAW_Images | 11_RGB_raw.png |
| Sample 11 | 0.11 | 14x14 | 1200 | 3 | 5 | 14 | 431.45 | 950 | 950 | 950 | 10 | 129 | 106 | 18.3 | 10 | RAW_Images | 12_RGB_raw.png |
| Sample 12 | 0.11 | 7x11 | 1200 | 3 | 5 | 14 | 431.45 | 950 | 950 | 950 | 10 | 129 | 106 | 18.3 | 10 | RAW_Images | 13_RGB_raw.png |
| Sample 13 | 0.11 | 10x12 | 1200 | 3 | 5 | 14 | 431.45 | 950 | 950 | 950 | 10 | 128 | 106 | 18.1 | 10 | RAW_Images | 14_RGB_raw.png |
| Sample 14 | 0.11 | 12x14 | 1200 | 3 | 5 | 14 | 431.45 | 950 | 950 | 950 | 10 | 128 | 106 | 18.1 | 10 | RAW_Images | 15_RGB_raw.png |
| Sample 14 | 0.11 | 11x12 | 1200 | 3 | 5 | 14 | 431.45 | 950 | 950 | 950 | 10 | 128 | 106 | 18.1 | 10 | RAW_Images | 16_RGB_raw.png |
| Sample 15 | 0.11 | 16x17 | 1200 | 3 | 5 | 14 | 431.45 | 950 | 950 | 950 | 10 | 129 | 106 | 18 | 10 | RAW_Images | 17_RGB_raw.png |
| Sample 16 | 0.12 | 13x13 | 1200 | 3 | 5 | 14 | 431.45 | 950 | 950 | 950 | 10 | 130 | 104 | 20 | 10 | RAW_Images | 18_RGB_raw.png |
| Sample 17 | 0.11 | 11x15 | 1200 | 3 | 5 | 14 | 431.45 | 950 | 950 | 950 | 10 | 129 | 106 | 18.7 | 10 | RAW_Images | 19_RGB_raw.png |
| Sample 18 | 0.11 | 23x8 | 1200 | 3 | 5 | 14 | 431.45 | 950 | 950 | 950 | 10 | 129 | 106 | 18.7 | 10 | RAW_Images | 20_RGB_raw.png |
| Sample 19 | 0.11 | 7x10 | 1200 | 3 | 5 | 14 | 431.45 | 950 | 950 | 950 | 10 | 130 | 104 | 20 | 10 | RAW_Images | 21_RGB_raw.png |
| Sample 19 | 0.11 | 10x18 | 1200 | 3 | 5 | 14 | 431.45 | 950 | 950 | 950 | 10 | 130 | 104 | 20 | 10 | RAW_Images | 22_RGB_raw.png |
| Sample 20 | 0.11 | 16x15 | 1200 | 3 | 5 | 14 | 431.45 | 950 | 950 | 950 | 10 | 130 | 104 | 20 | 10 | RAW_Images | 23_RGB_raw.png |
| Sample 01 | 0.11 | 16x16 | 1200 | 3 | 5 | 14 | 431.45 | 950 | 950 | 950 | 10 | 129 | 107 | 18.2 | 10 | BASIC_Images | 01_RGB_basic.png |
| Sample 02 | 0.11 | 10x11 | 1200 | 3 | 5 | 14 | 431.45 | 950 | 950 | 950 | 10 | 129 | 107 | 18.2 | 10 | BASIC_Images | 02_RGB_basic.png |
| Sample 02 | 0.11 | 12x12 | 1200 | 3 | 5 | 14 | 431.45 | 950 | 950 | 950 | 10 | 129 | 107 | 18.2 | 10 | BASIC_Images | 03_RGB_basic.png |
| Sample 03 | 0.11 | 10x11 | 1200 | 3 | 5 | 14 | 431.45 | 950 | 950 | 950 | 10 | 128 | 107 | 17.8 | 10 | BASIC_Images | 04_RGB_basic.png |
| Sample 04 | 0.11 | 14x16 | 1200 | 3 | 5 | 14 | 431.45 | 950 | 950 | 950 | 10 | 128 | 107 | 17.8 | 10 | BASIC_Images | 05_RGB_basic.png |
| Sample 05 | 0.11 | 8x14 | 1200 | 3 | 5 | 14 | 431.45 | 950 | 950 | 950 | 10 | 129 | 106 | 18.5 | 10 | BASIC_Images | 06_RGB_basic.png |
| Sample 06 | 0.11 | 16x16 | 1200 | 3 | 5 | 14 | 431.45 | 950 | 950 | 950 | 10 | 129 | 106 | 18.5 | 10 | BASIC_Images | 07_RGB_basic.png |
| Sample 07 | 0.11 | 14x10 | 1200 | 3 | 5 | 14 | 431.45 | 950 | 950 | 950 | 10 | 129 | 107 | 18.1 | 10 | BASIC_Images | 08_RGB_basic.png |
| Sample 08 | 0.11 | 14x12 | 1200 | 3 | 5 | 14 | 431.45 | 950 | 950 | 950 | 10 | 129 | 107 | 18.1 | 10 | BASIC_Images | 09_RGB_basic.png |
| Sample 09 | 0.11 | 15x16 | 1200 | 3 | 5 | 14 | 431.45 | 950 | 950 | 950 | 10 | 129 | 107 | 17.7 | 10 | BASIC_Images | 10_RGB_basic.png |
| Sample 10 | 0.11 | 15x14 | 1200 | 3 | 5 | 14 | 431.45 | 950 | 950 | 950 | 10 | 129 | 107 | 17.7 | 10 | BASIC_Images | 11_RGB_basic.png |
| Sample 11 | 0.11 | 14x14 | 1200 | 3 | 5 | 14 | 431.45 | 950 | 950 | 950 | 10 | 129 | 106 | 18.3 | 10 | BASIC_Images | 12_RGB_basic.png |
| Sample 12 | 0.11 | 7x11 | 1200 | 3 | 5 | 14 | 431.45 | 950 | 950 | 950 | 10 | 129 | 106 | 18.3 | 10 | BASIC_Images | 13_RGB_basic.png |
| Sample 13 | 0.11 | 10x12 | 1200 | 3 | 5 | 14 | 431.45 | 950 | 950 | 950 | 10 | 128 | 106 | 18.1 | 10 | BASIC_Images | 14_RGB_basic.png |
| Sample 14 | 0.11 | 12x14 | 1200 | 3 | 5 | 14 | 431.45 | 950 | 950 | 950 | 10 | 128 | 106 | 18.1 | 10 | BASIC_Images | 15_RGB_basic.png |
| Sample 14 | 0.11 | 11x12 | 1200 | 3 | 5 | 14 | 431.45 | 950 | 950 | 950 | 10 | 128 | 106 | 18.1 | 10 | BASIC_Images | 16_RGB_basic.png |
| Sample 15 | 0.11 | 16x17 | 1200 | 3 | 5 | 14 | 431.45 | 950 | 950 | 950 | 10 | 129 | 106 | 18 | 10 | BASIC_Images | 17_RGB_basic.png |
| Sample 16 | 0.12 | 13x13 | 1200 | 3 | 5 | 14 | 431.45 | 950 | 950 | 950 | 10 | 130 | 104 | 20 | 10 | BASIC_Images | 18_RGB_basic.png |
| Sample 17 | 0.11 | 11x15 | 1200 | 3 | 5 | 14 | 431.45 | 950 | 950 | 950 | 10 | 129 | 106 | 18.7 | 10 | BASIC_Images | 19_RGB_basic.png |
| Sample 18 | 0.11 | 23x8 | 1200 | 3 | 5 | 14 | 431.45 | 950 | 950 | 950 | 10 | 129 | 106 | 18.7 | 10 | BASIC_Images | 20_RGB_basic.png |
| Sample 19 | 0.11 | 7x10 | 1200 | 3 | 5 | 14 | 431.45 | 950 | 950 | 950 | 10 | 130 | 104 | 20 | 10 | BASIC_Images | 21_RGB_basic.png |
| Sample 19 | 0.11 | 10x18 | 1200 | 3 | 5 | 14 | 431.45 | 950 | 950 | 950 | 10 | 130 | 104 | 20 | 10 | BASIC_Images | 22_RGB_basic.png |
| Sample 20 | 0.11 | 16x15 | 1200 | 3 | 5 | 14 | 431.45 | 950 | 950 | 950 | 10 | 130 | 104 | 20 | 10 | BASIC_Images | 23_RGB_basic.png |